# Isolated and Ensemble Audio Preprocessing Methods for Detecting Adversarial Examples against Automatic Speech Recognition

**Krishan Rajaratnam** [1], **Kunal Shah** [2], **Jugal Kalita** [3]

[1] The College, University of Chicago, Chicago, USA
[2] College of Liberal Arts and Sciences, University of Florida, Gainesville, USA
[3] Department of Computer Science, University of Colorado, Colorado Springs, USA

[1] krajaratnam@uchicago.edu, [2] kshah1997@ufl.edu, [3] jkalita@uccs.edu

**Abstract**

An adversarial attack is an exploitative process in which minute alterations are made to natural inputs, causing the inputs to be misclassified by neural models. In the field of speech recognition, this has become an issue of increasing significance. Although adversarial attacks were originally introduced in computer vision, they have since infiltrated the realm of speech recognition. In 2017, a genetic attack was shown to be quite potent against the Speech Commands Model. Limited-vocabulary speech classifiers, such as the Speech Commands Model, are used in a variety of applications, particularly in telephony; as such, adversarial examples produced by this attack pose as a major security threat. This paper explores various methods of detecting these adversarial examples with combinations of audio preprocessing. One particular combined defense incorporating compressions, speech coding, filtering, and audio panning was shown to be quite effective against the attack on the Speech Commands Model, detecting audio adversarial examples with 93.5% precision and 91.2% recall.

## 1. Introduction

Due to the widespread and growing use of neural networks for various tasks, it is imperative that these models be robust and secure while remaining generally usable. Although these models are quite powerful and are well-suited for a variety of tasks, they are not without their imperfections. First applied against computer vision models [1], adversarial attacks exploit the flaws of neural networks by making perceptibly insignificant changes to a source to produce adversarial examples, with the purpose of causing the neural network to misclassify the example. These attacks can be quite potent and have caused misclassification rates of above 90% in image classifiers [2]. Because of their exploitative nature, adversarial attacks can be quite difficult to defend against without sacrificing general model usability or accuracy.

The use of adversarial attacks is not restricted to the field of image recognition. Modern speech recognition has become increasingly reliant on end-to-end neural models, which are able to largely outperform traditional models that rely heavily on signal processing and hidden Markov models. These sophisticated neural models may be state-of-the art, but are also more susceptible to attack by adversarial examples. Recent work has shown that two speech recognition models, a convolutional neural network (CNN) model trained on the Speech Commands dataset [3] and Mozilla's implementation of the DeepSpeech end-to-end model [4], are vulnerable to adversarial attacks. Two separate attacks on the two models were able to generate extremely potent adversarial examples, capable of inducing a misclassification rate of up to 100%. This trend threatens the current reliability of deep learning models within the field of speech recognition. As such, there is a crucial need for defensive methods that can be employed to evade audio adversarial attacks.

## 2. Related Work

The attack against the limited-vocabulary Speech Commands model detailed by Alzantot et al. [3] shows particular relevance within the field of telephony, as it could be applied to maliciously manipulate the limited-vocabulary speech classifiers used for automated attendants. During this attack, adversarial examples created by a gradient-free genetic algorithm allow the attack to penetrate layers of non-differential preprocessing, which is commonly used in automatic speech recognition.

### 2.1 Audio Preprocessing Defenses

Recent work in computer vision has shown that preprocessing methods, such as JPEG and JPEG2000 image compression [5], resizing [6], and pixel deflection [7] are capable of defending against adversarial attacks with varying degrees of success. Preprocessing defenses have also been employed in speech recognition to mitigate adversarial examples. Yang et al. [8] achieved a high rate of success with the use of local smoothing, down sampling, and quantization in an attempt to neutralize adversarial examples produced by the attack of Alzantot et al. Quantizing with $q = 256$, Yang et al. achieved their best result of mitigating

(i.e. retrieving the original label of) 63.8% of the adversarial examples. Quantization causes various amplitudes of sampled data to be rounded to the nearest integer multiple of the *q* value; this allows adversarial perturbations with small amplitudes to become disrupted.

Work has also been done in employing audio compression, Hertz shifting, noise reduction, and low-pass filtering [9], to defend against Carlini and Wagner's attack [4] on DeepSpeech. The results of [9] suggest that the most promising preprocessing method was low-pass filtering, with which the authors were able to retrieve the original label of 90.11% of Carlini and Wagner's adversarial examples. By utilizing low-pass filtering, a selected range of higher frequencies are eliminated, preserving a lower band of frequencies in which human speech is located. If a significant portion of the of the adversarial perturbation is found within the discarded higher frequencies, the attack can be disrupted. Although this work was able to largely neutralize the threat of adversarial examples against DeepSpeech, this came at a noticeable cost to general model accuracy.

2.2 Speech Coding

Although the results of [9] imply that low-pass filtering outclasses audio compression as a preprocessing defense, this work only explored two standards of audio coding: Advanced Audio Coding (AAC) and MP3. Though those two compression standards enjoy widespread popularity, they are not necessarily adequately equipped in defending against targeted adversarial examples on speech recognition. For the purposes of teleconferencing and VoIP, speech codecs such as Speex [10] and Opus [11] are primarily used due to their ability to preserve the quality of human speech, even through imperfect conditions and lower bitrates.

In 2002, Valin [10] began the Speex project with intent on providing "a free codec for free speech." Further development allowed Speex to grow in popularity, becoming adopted by well-known, practical VoIP applications such as TeamSpeak[1] and Twinkle[2]. Speex codec is built upon the Code Excited Linear Prediction (CELP) algorithm [12], which models the vocal tract using a linear prediction model capable of minimizing differences of the uncompressed source within a "perceptually weighted domain." The minimization is achieved by applying the following weighting filter to the raw input:

$$W(z) = \frac{A(z/\gamma_1)}{A(z/\gamma_2)} \qquad (1)$$

---

[1] http://teamspeak.com/en/features/overview
[2] http://www.linuxlinks.com/Twinkle/

where $A$ is a linear prediction filter with $\gamma_1$ and $\gamma_2$ managing the filter shape. This filter allows for various levels of noise at different frequencies and has proven to be useful for neutralizing adversarial perturbations whilst maintaining the quality of human speech. In addition, Speex also includes numerous features, such as voice activity detection, denoising, and support of various bandwidths. As this compression seems to resemble audio preprocessing methods proven capable of effectively mitigating adversarial examples, it seems better suited than MP3 or AAC compression for the task of defending against adversarial attacks.

The Opus codec, which is used by the highly popular proprietary VoIP application Discord[3], is a modern successor to the Speex codec [11]. By combining the CELP algorithm with SILK, a linear predictive coding algorithm developed by Skype Technologies in 2009[4], it is considered an improved and more advanced version of Speex; the application of Opus compression for defending against adversarial examples is therefore worth testing.

2.3 Ensemble Detection

Preprocessing defenses against adversarial examples can only be effective and practical if they are able to mitigate adversarial examples without greatly compromising general model accuracy. A viable form of preprocessing would disrupt the predictions of adversarial examples more than it would disrupt the predictions of benign examples. In particular, there should ideally be a small difference between the output vectors produced by passing the raw input and preprocessed input through a neural network when the input is benign, but that same difference should be much larger if the input is adversarial. This core idea can be used to apply preprocessing methods to detect adversarial examples, rather than simply mitigating or neutralizing perturbations.

Within the field of computer vision, ensembles of preprocessing methods have been used for detecting adversarial examples. Xu et al. [13] proposed the feature squeezing method for detecting adversarial examples. This method combines smaller "squeezing" methods into an ensemble, and calculates an $L_1$ score from of the maximum $L_1$ distance between any pair of output probability vectors produced by passing the raw and squeezed inputs through a deep neural network (DNN). Using feature squeezing, Xu et al. were able to consistently detect over 80% of adversarial examples produced from a variety of attacks.

---

[3] http://discordapp.com/features
[4] http://www.h-online.com/open/news/item/Skype-publishes-SILK-audio-codec-source-code-955264.html

## 3. Methods and Evaluation

The aim of this research can be divided into two parts: using the individual methods of preprocessing independently to detect adversarial examples, and examining various methods of combining the preprocessing detectors together as ensemble detection methods. The adversarial examples are generated using the genetic attack described by Alzantot et al. against the pre-trained Speech Commands model [3].

### 3.1 Speech Commands Dataset and Model

The Speech Commands dataset was released in 2017 and contains 105,829 labeled utterances of 32 words from 2,618 speakers [14]. As a light-weight model, Speech Commands is based on a keyword-spotting convolutional network (CNN) [15] that is capable of achieving 90% classification accuracy on this dataset. For the purposes of this research, a unique subset of 30,799 labeled utterances of 10 words are used in order to maintain consistency with previous research pertaining to the adversarial examples of Alzantot, et al. From this subset, 20 adversarial examples are generated for each nontrivial source-target word pair for 1800 total examples. Each example is produced with a maximum of 500 iterations.

### 3.2 Preprocessing Defenses

A simple method for using preprocessing to detect adversarial examples is by checking to see if the prediction produced by the model changes if the input is preprocessed; if the model's prediction of the raw input does not match the prediction of the preprocessed input, it is declared adversarial. The following preprocessing methods are used in isolation for detecting adversarial examples:

- MP3 Compression,
- AAC Compression,
- Speex Compression,
- Opus Compression,
- Band-pass Filtering, and
- Audio Panning and Lengthening.

While the MP3 and AAC compressions correspond directly to preprocessing defenses in related work described in Section 2.1, the other defenses listed above have not yet been directly tested against audio adversarial examples. The band-pass filter defense builds off of the low-pass filter of [9] by combining it with a high-pass filter in order to deter additional adversarial perturbations outside of the frequency range for natural human speech. Audio panning is a form of preprocessing frequently used in audio mixing that distributes a signal across stereophonic channels, distorting channel volumes to mimic the perception of audio coming from an off-centered position. The audio panning and lengthening defense lengthens audio by 1% in addition to panning to increase the spatial distortion of adversarial perturbations in the signal.

3.3 Ensemble Detection Methods

Individual preprocessing methods as isolated defenses can successfully fend off certain adversarial attacks. However, attacks aware of the preprocessing defenses are capable of optimizing to become more robust [4]. As such, the use of any one preprocessing method alone for detecting adversarial examples would prove to be insufficient and render the model increasingly susceptible to more advanced attacks. Therefore, a combined deployment of preprocessing methods, or an ensemble, may be able to provide better security with a more complex defense.

The preprocessing detection methods described in Section 3.2 can be combined in a variety of configurations. The ensemble detection methods explored in this research are discussed below.

3.3.1 Majority Voting Ensemble

The simplest method of combining the preprocessing methods together would be by assigning each preprocessing method a vote, and declaring an audio signal as adversarial if a majority of the ensemble declares the signal adversarial. As there are six preprocessing methods that are combined into an ensemble, ties with this discrete voting scheme are possible. To err on the side of security, this procedure will declare a signal as adversarial in the event of a tie.

### 3.3.2 Learned Threshold Voting Ensemble

The majority voting ensemble declares an audio signal as adversarial if there are at least three votes in favor of it being adversarial. This threshold for deciding how many votes are needed to declare an audio signal as adversarial is arbitrary, and can adapt to different circumstances. A low threshold would result in a high recall in detecting adversarial examples, but would sacrifice precision. A high threshold would result in a lower recall in detecting adversarial examples, but would yield a higher precision. This ensemble method experiments with using various voting thresholds for detecting adversarial examples on a labeled training set, and chooses the threshold that results in the best precision and recall. To balance both precision and recall, $F_1$ scores are used for selecting the best threshold, although in practice, one could adjust the F-measure to reflect one's attitude on the relative importances of precision and recall.

### 3.3.3 $L_1$ Scoring

The previously discussed ensemble voting methods are relatively simple, as they simply examine the model's discrete prediction of the raw and preprocessed inputs for each preprocessing method. Additionally, the voting methods above are indiscriminate and treat each member of the ensemble equally. A more nuanced approach for measuring the differences in predictions between raw and preprocessed inputs is by $L_1$ scoring the different output logit vectors, similar to how Xu et al. integrated the multiple squeezing methods in their feature squeezing defense. In this method, an ideal threshold $L_1$ score is learned from training data by finding the threshold of maximum information gain, and test examples that surpass this threshold are declared adversarial. This method uses the maximum $L_1$ distance to calculate the score, implicitly assigning more importance to preprocessing methods that produce output vectors that are highly different than the output vectors produced by predicting raw signal. As such, this method would theoretically be more sensitive in detecting adversarial examples, but it may also be quite aggressive in declaring signals as adversarial at the risk of falsely declaring benign examples as adversarial.

### 3.3.4 Tree-based Classification Algorithms

The above ensemble methods discard information of the class-specific variation in the output vector for each preprocessing method, relative to the raw input. In order to preserve this information, a multi-dimensional vector can be used, with each dimension accounting for the output vector variation for that class. For the tree-based detection methods discussed in this research, a multi-dimensional vector composed of the summed absolute class-specific differences between the raw input's resultant probability vector and the preprocessed input's resultant probability vector over each method of preprocessing. In particular, the *i*th dimension of this summed absolute difference (SAD) vector *S* is calculated as follows:

$$S_i = \sum_{p \in P} |r_i - p_i| \qquad (2)$$

where *P* corresponds to the set of output probability vectors yielded by the methods of preprocessing in the ensemble, and *r* corresponds to the output probability vector produced by passing the raw signal through the Speech Commands model without any preprocessing.

This vector will preserve information about class-specific variation between the predictions, and will reduce the number of features of the vector inputted to the tree-based classifier down to 12 (which is the same as the number of classes). Considering the relatively small training dataset size (which is discussed in Section 3.4), having less features for tree-based classification may improve performance. However, the 84-dimensional vector formed by simply concatenating each output probability vector together would preserve the most amount of information. As such, the use of this concatenated probability (CP) vector for tree-based classification is also tested, even if the dataset isn't large enough for the classification algorithms to effectively handle that large of a vector.

Decision tree-based classification algorithms are well-suited for classifying vectors of features into discrete classes. In this research, three tree-based classification algorithms are employed for using vectors of summed absolute differences for detecting adversarial examples: random forest classification, adaptive boosting, and extreme gradient boosting. Random forest classification functions by constructing many decision trees in an attempt to stave off the possibility of over-fitting. Adaptive boosting and extreme gradient boosting are gradient boosting algorithms which function by building an ensemble of weak learners in a stage-wise fashion. Each of these tree-based algorithms are used twice in this research: once for using SAD vectors for classification and once for using CP vectors for classification. These tree-based algorithms have had quite high success in applied problems, are possibly well-suited for detecting adversarial examples.

3.4 Evaluation

The aforementioned detection methods are evaluated based on their precision and recall in detecting adversarial examples. As the simple preprocessing detection methods discussed in Section 3.2 require no training, the precision and recall measurements are calculated based off of their performances on the full set of 1,800 generated adversarial examples and 1,800 randomly selected benign examples. Many of the ensemble detection methods, however, do train and adapt based off what is seen in training data, so precision and recall measurements for these detection methods are calculated based off of their performance on a subset of only 900 adversarial examples and 900 benign examples; the 900 other adversarial and benign examples are used as a training dataset.

Within the context of defending against adversarial attacks, there seems to be an implicit tradeoff between security and general model accuracy. Although it is important to have a high recall in detecting adversarial examples for the sake of security, a low precision in detection would cause the model to decline in usability. This research takes the stance of both security and general model accuracy being equally important. To reflect this attitude, $F_1$ scores are used to combine the precision and recall measurements with equal consideration.

## 4. Results

The results of the individual preprocessing detection methods described in Section 3.2 are summarized in Table 1. Measurements indicate that all of the methods are capable of detecting adversarial examples produced by the attack with varying rates of success. The results are consistent with the findings of [9] in that MP3 compression performs adequately at best when compared with the other methods. AAC and Opus compression perform notably better, but are not able to achieve as high of a recall as Speex compression (which also yields the highest $F_1$ score).

Although the use of band-pass filtering for detecting adversarial examples is extremely precise, it yields a remarkably low recall, which suggests it is a bit too passive with its declaration of adversariality.

As many of these preprocessing methods distort audio signals in fundamentally different ways, the overall high precision (and lower recall) measurements of each of the individual preprocessing suggest that some of the ensemble methods may be more effective in detecting adversarial examples.

**Table 1:** *Precision, recall, and $F_1$ values for isolated preprocessing methods in detecting adversarial examples.*

| Preprocessing Method | Precision | Recall | $F_1$ Score |
|---|---|---|---|
| MP3 Compression | 93.7% | 70.7% | 0.806 |
| AAC Compression | 95.0% | 81.2% | 0.876 |
| Band-Pass Filtering | **97.3%** | 40.6% | 0.573 |
| Audio Panning & Lengthening | 95.8% | 82.4% | 0.886 |
| Opus Compression | 94.5% | 81.8% | 0.877 |
| Speex Compression | 93.7% | **88.5%** | **0.910** |

**Table 2:** *Precision, recall, and $F_1$ values for ensemble detection methods in detecting adversarial examples.*

| Ensemble Detection Method | Precision | Recall | $F_1$ Score |
|---|---|---|---|
| Majority Voting | **96.1%** | 88.1% | 0.919 |
| Learned Threshold Voting | 93.5% | 91.2% | **0.924** |
| $L_1$ Scoring | 76.9% | 92.4% | 0.840 |
| Random Forest Classification (SAD Vector) | 79.3% | 87.0% | 0.830 |
| Random Forest Classification (CP Vector) | 86.7% | **94.4%** | 0.904 |
| Adaptive Boosting (SAD Vector) | 83.5% | 81.8% | 0.827 |
| Adaptive Boosting (CP Vector) | 86.7% | 93.0% | 0.898 |
| Extreme Gradient Boosting (SAD Vector) | 83.0% | 84.2% | 0.836 |
| Extreme Gradient Boosting (CP Vector) | 88.3% | **94.4%** | 0.913 |

The results of the ensemble detection methods described in Section 3.3 are summarized in Table 2. The voting methods performed quite well and achieved the two highest $F_1$ scores of all the methods discussed in this paper. This may be attributed to the high precisions and low recalls of the individual preprocessing methods described in Table 1; the relatively strict voting threshold of votes needed for an adversarial declaration capitalizes on the high precision of each of the methods and is able to increase recall. The majority voting method especially benefited from the high precisions of its constituents and yielded an extremely high precision of 96.1%. The Learned Threshold Voting method was able to learn a lower voting threshold of only two votes needed for an adversarial declaration. As such, this method was able to yield a notably higher recall than what was achieved through majority voting, but at a noticeable cost to precision. As the Learned Threshold Voting method still retained a fairly high precision, it achieved the overall highest $F_1$ score of any of the other preprocessing methods. The recall values for detecting adversarial examples using the Learned Threshold Voting method are detailed in Figure 1.

The $L_1$ Scoring method was able to achieve higher recall than either of the two voting methods, perhaps due to its aggressive nature. However, this was achieved at the cost of precision, which evidently lowered the $F_1$ score.

Although tree-based classification algorithms can be quite powerful in a variety of situations, the tree-based methods were not able to perform as well as the voting methods in detecting adversarial examples using SAD vectors. This may be because the SAD vectors fed into the tree algorithms discarded important voter-specific information. In particular, the vector of summed absolute differences effectively anonymizes the voters in the ensemble; it inherently considers each member of the ensemble equally.

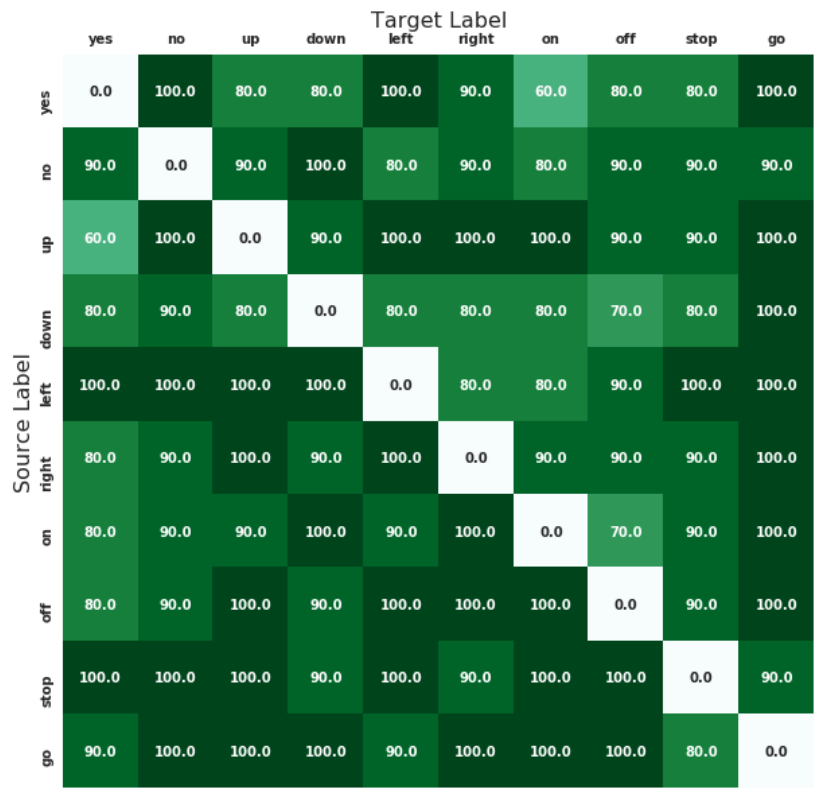

**Figure 1:** *A heat map detailing the rates (in percent) of detecting targeted adversarial examples with the Learned Threshold Voting method. The diagonal of zeroes correspond to trivial source-target pairs for which no adversarial examples were generated.*

This discarded information proved to be quite crucial for effectively detecting adversarial examples, as the tree-based classification methods performed significantly better with CP vectors (which are highly conservative). In particular, the extreme gradient boosting and adaptive boosting classification algorithms were able to yield the highest recall values for detecting adversarial examples out of all of the detection methods discussed in this research. Considering that the tree-based classification methods performed significantly better with the voter-specific information available in the CP vector, it is worth noting that the Learned Threshold Voting method, which yielded a higher $F_1$ score than any tree-based classification method, does not use voter-specific information; each vote carries equal weight towards breaking the learned threshold. As such, it may be possible that the tree-based classification methods outperform the Learned Threshold Voting method on larger datasets, as it could be that this training dataset was not sufficiently large enough for learning how to optimally use an 84-dimensional vector for classification. However, given the heavy reliance of training data that the tree-based classification methods exhibit, they are likely not as well-suited for flexibly handling different types of attacks as the voting methods.

As the Learned Threshold Voting method performed better over all other detection methods discussed in this paper, it can be helpful to examine the adversarial examples that remain undetected by this method and the benign signals that get incorrectly flagged as adversarial.

One method of analyzing the adversarial examples is by examining the average frequency level throughout the signal. Since we are able to recover the original, clean source for each adversarial example, we can examine the difference between the average frequencies of each adversarial example and its clean source. The means and standard deviations of this difference for undetected and detected adversarial examples are depicted in Table 3.

**Table 3:** *Mean and Standard Deviation of Differences of Average Frequencies for undetected and detected adversarial examples.*

|  | Mean Difference (Hz) | Standard Deviation (Hz) |
|---|---|---|
| Undetected Adversarial Examples | 988 | 464 |
| Detected Adversarial Examples | 1176 | 590 |

Upon performing a Student's t-test, the difference in undetected adversarial examples was found to be less than the difference in detected adversarial examples with 99% statistical significance. This suggests that the frequencies of adversarial perturbations in the undetected adversarial examples were concentrated at lower frequencies than those of detected adversarial examples. As human speech is found within these lower frequencies, it is much more difficult to disrupt or detect these adversarial examples without distorting the speech in the signal. This may explain why these adversarial examples remained undetected under the Learned Threshold Voting method. As a side effect, these undetected adversarial examples with significant perturbation in the lower frequency bands would theoretically be more perceptibly noisy to humans, as the physiology of the inner ear is fine-tuned for picking up auditory information at these frequencies [16].

Benign examples that were falsely detected as adversarial also exhibited an interesting property. In particular, the Speech Commands model achieved a classification accuracy of 92.7% on raw benign examples that were not detected as adversarial, but that accuracy fell to 40.4% for raw benign examples that were falsely detected as adversarial. The relatively low classification accuracy for the benign examples flagged as adversarial suggests that even benign examples that are classified incorrectly by the model exhibits some volatility of outputted predictions upon preprocessing, similar to adversarial examples. For reference, the model achieved 90.3% classification accuracy in general over all benign examples.

## 5. Conclusion and Future Work

Although the results of this research suggest that ensembles of audio preprocessing can be highly effective for detecting adversarial examples, it is important to note the drawbacks of the defenses discussed in this paper. An analysis of adversarial examples that went undetected by the Learned Voting Threshold method implied that those examples had more adversarial perturbations in the lower frequency bands than the adversarial examples that were detected. This suggests that attacks can optimize adversarial examples robust to the Learned Threshold Voting method by concentrating adversarial perturbations within the frequency range of human speech.

Although using ensemble detection methods may provide marginal security over using isolated preprocessing detection methods, recent work has shown that adaptive attacks on image classifiers are able to bypass ensembles of weak defenses [17], including the feature squeezing ensemble of Xu, et al.; this work could be applied to attack speech recognition models. Future work can be done in investigating stronger ensembles for detecting audio adversarial examples and other defenses that can withstand adaptive attacks. Considering the faceted usefulness of Speex compression for detecting adversarial examples, perhaps further investigation into speech coding for defending against adversarial attacks is warranted.

Nevertheless, this paper demonstrated that methods of audio preprocessing can be used to detect adversarial examples produced by the attack of Alzantot et al. on Speech Commands. Additionally this paper examined the effectiveness of various ensembles of audio preprocessing detection methods for defending against adversarial examples. While these detection methods may not be extremely effective against more adaptive attacks, this research aimed ultimately to further discussion of defenses against adversarial examples within the audio domain: a field in desperate need of more literature.

**Acknowledgments**

We are thankful to the reviewers for helpful criticism, and the UCCS LINC and VAST labs for general support. This work is supported by the National Science Foundation under Grant No. 1659788. Any opinions, findings, and conclusions or recommendations expressed in the material are those of the author(s) and do not necessarily represent the views of the National Science Foundation.